\documentclass[aps,prd,twocolumn,showpacs,nofootinbib]{revtex4}

\usepackage{epsfig,amssymb,euscript}
\usepackage{amsmath,amscd}
\newcommand{\be}{\begin{eqnarray}}
\newcommand{\ee}{\end{eqnarray}}
\newcommand{\bea}{\begin{eqnarray}}
\newcommand{\eea}{\end{eqnarray}}
\newcommand{\ba}{\begin{array}}
\newcommand{\ea}{\end{array}}

\newcommand{\nn}{\nonumber \\}

\newcommand{\p}[1]{(\ref{#1})}

\newcommand{\bR}{\mathbb{R}}

\newcommand{\bchi}{{\mbox{\boldmath $\chi$}}}
\newcommand{\bomega}{{\mbox{\boldmath $\omega$}}}

\begin{document}

\title{General Concentric Black Rings}

\author{Jerome P. Gauntlett}

\affiliation{Perimeter Institute for Theoretical
Physics\\ Waterloo, ON, N2J 2W9, Canada
\\
\\ (On leave from: Blackett Laboratory, Imperial
  College, London, SW7 2AZ, U.K.)}

\author{Jan B. Gutowski}

\affiliation{Mathematical Institute, Oxford University,
\\Oxford, OX1 3LB, U.K.}

\date{\today}

\begin{abstract}
\noindent Supersymmetric black ring solutions of five dimensional
supergravity coupled to an arbitrary number of vector
multiplets are constructed.
The solutions are asymptotically flat and
describe configurations of concentric black rings
which have regular horizons with topology $S^1 \times S^2$ and
no closed time-like curves at the horizons.

\end{abstract}

%\keywords{}

\maketitle

\section{Introduction}
An interesting development in black hole studies is the discovery
of black rings in five-dimensions.
These are asymptotically flat black hole solutions with an
event horizon of topology $S^1\times S^2$, rather
than the more familiar $S^3$. They
were first discovered in pure Einstein gravity \cite{harveyrng:2002}
and were further generalised in
\cite{elvang:2003yy,elvang:2003mj,emparan:2004}.
More recently, supersymmetric black rings have been found
in D=5 minimal supergravity \cite{eemr} (see \cite{bena} for
earlier related work) and these have been generalised to
give supersymmetric multi-concentric black ring
solutions of the same theory \cite{jgconcentric}.
Furthermore, the construction in \cite{jgconcentric} also allows
for the possibility of a rotating black
hole to sit at the common centre of the rings.

The black ring solutions all have non-vanishing rotation which supports
them from collapsing.
A single supersymmetric black ring has two angular momenta which are
necessarily different, and
hence the single black ring is distinguished by its asymptotic charges
from the supersymmetric rotating black hole solution of
\cite{tseytlin:1996,Breckenridge:1996is}, (see also \cite{gauntlett:1998}).
An interesting feature of the concentric ring solutions
\cite{jgconcentric} is that they can carry the same
conserved charges as the rotating black hole and with an entropy
that can be smaller than, equal to
or greater than that of the black hole. This shows
that the conserved charges of the multi-ring configurations in themselves
are not sufficient to distinguish the
microstates that presumably account for the entropy
of the rings after embedding them in string theory.

In this paper we will further generalize the results of \cite{eemr}
and construct supersymmetric black
ring solutions of D = 5 minimal supergravity coupled to an
arbitrary number of vector multiplets.
We will also generalize the results of \cite{jgconcentric}
by constructing multi-concentric black ring
solutions of this more general class of supergravity theories.
Recall that these theories can arise, for example,
as part of the low-energy effective action
of M-theory reduced on a Calabi-Yau 3-fold \cite{cad:1995}.
The scalar field in each vector multiplet then corresponds to
a coordinate on the K\"ahler moduli space of the
Calabi-Yau (excluding the overall volume, which appears
in a hypermultiplet). Note that general static and
rotating black hole solutions of this theory have
been found in \cite{sabra:97b} and
\cite{Chamseddine:1998yv}, respectively.

An important feature of black hole solutions of these theories is
that the near horizon geometry is essentially
independent of the asymptotic
values of the scalar fields at infinity (this was first discovered in
a four-dimensional context in
\cite{Ferrara:1995ih} and further explored in
\cite{Strominger:1996sh,Ferrara:1996dd}. A discussion of the
five-dimensional case is also given
in \cite{Ferrara:1996um,wong:96,Chou:1997ba}.)
In the case that the supergravity theory is obtained
from a compactification of a Calabi-Yau
manifold, the asymptotic values of
the scalars specify particular moduli of the Calabi-Yau.
The near horizon limit of the black hole geometry is therefore
independent of these moduli.

In order to have a statistical interpretation, it is expected that
the black hole entropy should not depend on adiabatic changes of
the environment and hence not on the scalar moduli \cite{larsen:1995}.
Thus, one  motivation for the present work is to see
if a similar phenomenon happens for general supersymmetric black rings.
For a single black ring carrying multiple charges,
we shall see that it indeed does: the area of the horizon of the
ring depends on the scalar moduli but only via the difference of
the two conserved angular momentum. We also discuss the generalisation
of this statement to concentric rings.

The construction of the solutions will follow the
strategy of \cite{jgconcentric}.
We use the classification of the most general supersymmetric solutions
of D=5 gauged supergravity coupled to an arbitrary number of vector
multiplets \cite{gutharv1}, generalising
that of \cite{classb,Gauntlett:2003fk}.
Although the classification developed in \cite{gutharv1}
is for the gauged theory, we can easily extract the
appropriate results for the ungauged theory.
In particular we are only interested in
the ``time-like case", where the vector that can be constructed as a
bi-linear from the Killing spinor
is not everywhere null. In a neighbourhood where
this vector is time-like, we have a canonically defined
hyper-K\"ahler metric. Here, generalising a
similar analysis of \cite{classb}, we study the case when this base is
a Gibbons-Hawking space \cite{Gibbons:1979zt},
when the analysis simplifies considerably. With
this technology in hand, and the results of \cite{jgconcentric},
our construction of the general concentric rings is straightforward.

The plan of the rest of the paper is as follows. We first
summarise the general results of
the classification of supersymmetric solutions
in section 2, and then analyse
the special case of a Gibbons-Hawking base in section 3.
In section 4 we present our
new multi-concentric black ring solutions, which
once again can have an optional
rotating black hole sitting at the common centre.
In section 5 we present some further
details for the special case of the three-charge ``STU''  model.
This model can be obtained from the dimensional reduction
of D=11 supergravity on a six-torus.
It is therefore trivial to uplift our solutions to obtain
solutions of D=11 supergravity
and, after reduction and T-duality, solutions of type IIB supergravity
reduced on a five-torus. Section 6 briefly concludes.

\section{Supersymmetric solutions of ${\cal N}=1$ supergravity}

\subsection{${\cal N} = 1$ supergravity}

The action of ${\cal N}=1$ $D=5$ ungauged supergravity coupled to $n-1$
abelian vector multiplets is given by \cite{Gunaydin:1983bi}
\bea
 S &=& {1 \over 16 \pi G} \int \bigg( {}^5 R  -Q_{IJ} F^I
 \wedge *F^J -Q_{IJ} dX^I \wedge
 * dX^J
\nn
  &-&{1 \over 6} C_{IJK} F^I \wedge F^J \wedge A^K \bigg)
\eea
where we use a positive signature metric and the fermions have
been set to zero. $I,J,K$ take values $1 , \ldots ,n$ and $C_{IJK}$
are constants that are symmetric on $IJK$.
The $X^I$ are scalars which are constrained
via
\be
\label{eqn:conda}
{1 \over 6}C_{IJK} X^I X^J X^K=1 \ .
\ee
We may regard the $X^I$ as being functions of $n-1$
unconstrained scalars $\phi^a$. It is convenient to define
\be
X_I \equiv {1 \over 6}C_{IJK} X^J X^K
\ee
so that the condition ({\ref{eqn:conda}}) becomes
\be
X_I X^I =1 \ .
\ee
In addition, the coupling $Q_{IJ}$ depends on the scalars via
\be
Q_{IJ} = {9 \over 2} X_I X_J -{1 \over 2}C_{IJK} X^K \ .
\ee
In the special case that the scalars $X^I$ take
values in a symmetric space we have the important identity
\be
\label{eqn:jordan}
C_{IJK} C_{J' (LM}
C_{PQ) K'} \delta^{J J'} \delta^{K K'} = {4 \over 3} \delta_{I (L}
C_{MPQ)}.
\ee
In this case we have the relation:
\be\label{symid}
X^I=\frac{9}{2} C^{IJK} X_J X_K
\ee
where $C^{IJK}\equiv \delta^{II'}\delta^{JJ'}\delta^{KK'}C_{I'J'K'}$. The
constraints \p{eqn:jordan} are also sufficient to ensure that the matrix
$Q_{IJ}$ is invertible with an inverse $Q^{IJ}$ given by
\be\label{qinv}
Q^{IJ}=2 X^I X^J-6 C^{IJK}X_K.
\ee
Note that the equations of motion and supersymmetry transformations
for this theory (with opposite signature) can be found, for
example, in \cite{gutharv1}.

%The Einstein equation is given by
%\bea
%\label{eqn:ein}
%-{}^5 R_{\alpha \beta} +Q_{IJ} F^I{}_{\alpha \lambda}
%F^J{}_\beta{}^\lambda+Q_{IJ} \nabla_\alpha X^I \nabla_\beta X^J
%-{1 \over 6}g_{\alpha \beta} \left(Q_{IJ} F^I{}_{\mu \nu}
% F^{J \mu \nu} \right) =0
%\eea
%the gauge equations are
%\be
%\label{eqn:gauge}
%d \left(Q_{IJ} \star F^J \right)=-{1 \over 4}C_{IJK} F^J \wedge F^K \ ,
%\ee
%and the scalar equation can be written as
%\bea
%d \left(\star dX_I \right) - \left({1 \over 6} C_{MNI} -{1 \over
%2}X_I C_{MNJ} X^J \right) dX^M \wedge \star dX^N \nn
%+ \left( X_M X^P C_{NPI}-{1 \over 6}C_{MNI}-6 X_I X_M X_N+{1 \over 6}
%X_I C_{MNJ} X^J \right) F^M \wedge \star F^N=0 .
%\eea

By definition, a bosonic supersymmetric background
admits a Killing spinor
$\epsilon^a$. From this Killing spinor we can construct
tensors from spinor bi-linears, which can be used to classify the general
supersymmetric solutions of this theory. Building on the work
\cite{classb,Gauntlett:2003fk} this has been
carried out for the more general gauged theory
in \cite{gutharv1}. The relevant equations
for the ungauged theory, of interest
here, can be obtained from those in \cite{gutharv1} by simply
setting the gauge parameter $\chi$ to vanish (with some care).

There are two types of supersymmetric geometries, specified by whether
the vector $V$ that can be constructed from the Killing spinor is
null everywhere, the ``null case'', or not, the ``time-like case''.
Here, we will only be interested in the latter case. We also
note that in general $V$ is a Killing vector field that
generates a symmetry of the full solution.

\subsection{The time-like case}

In the time-like case we work in a neighbourhood ${\cal U}$ where
$V$ is a time-like Killing vector field.
Introduce coordinates $(t,x^m)$ such that
$V = \partial/\partial t$. The conditions for the existence
of time-like Killing spinors can then be summarised as follows.
The metric can be written locally as
\be
 \label{eqn:metric}
 ds^2=-f^2(dt+\omega)^2+f^{-1}ds^2(M_4)
\ee
where $M_4$ is an arbitrary four-dimensional
hyper-K\"ahler manifold, and
$f$ and $\omega$ are a scalar and a one-form on $M_4$, respectively.
We define
\be
 \label{eqn:e0def}
 e^0 = f (dt+\omega).
\ee
and choose the orientation of $M_4$ so that $e^0 \wedge \eta_4$ is
positively oriented in five dimensions, where $\eta_4$ is the volume
form of $M_4$.  We can split the two-form $d\omega$
into self-dual and anti-self-dual parts on $M_4$
and it is convenient to define:
\be
\label{eqn:rsp}
f d\omega\equiv G^{+}+G^{-}
\ee
The gauge field can then be written
\be
\label{eqn:rewritegaug}
F^I = d (X^I e^0) + \Theta^I \ ,
\ee
where $\Theta^I$ is a
self-dual two-form on $M_4$ satisfying
\be
\label{eqn:gpluscontr}
X_I \Theta^I = -{2 \over 3} G^+ \ .
\ee
These conditions are sufficient to ensure the existence of
a Killing spinor preserving 4 of the 8 supersymmetries.

We now consider the consequence of imposing the
Bianchi identities $dF^I=0$ and the Maxwell equations.
The Bianchi identities give
\be
\label{eqn:bianch}
d  \Theta^I=0 \ ,
\ee
so the $\Theta^I$ are harmonic self-dual two-forms on the base.
The Maxwell equations reduce to
\bea
\label{eqn:timegauge}
\nabla^2_{HK} \left( f^{-1} X_I \right) &=& {1 \over 6}C_{IJK}
(\Theta^J \ . \ \Theta^K) \ ,
\eea
where $\nabla^2_{HK}$ denotes the Laplacian
on the hyper-K\"ahler base $M_4$;
and contracting ({\ref{eqn:timegauge}}) with $X^I$ we obtain
\bea
\label{eqn:asympthet}
\nabla^2_{HK} f^{-1}  &=&
-{1 \over 3}Q_{IJ} \big[ (\Theta^I \ . \  \Theta^J)
+ 2 f^{-1}  (dX^I \ . \ dX^J) \big]
\nn
&+&{2 \over 3} (G^+ \ . \ G^+) \ ,
\eea
where we have used the convention that for
$p$-forms $\alpha$, $\beta$ on $B$, we set
\be
(\alpha \ . \  \beta) = {1 \over p!} \alpha_{m_1 \dots m_p}
\beta^{m_1 \dots m_p}\ .
\ee
The integrability conditions for the existence of a Killing spinor
guarantee that the Einstein equation and scalar equations of motion
are satisfied as a consequence of the above equations.

\section{Gibbons-Hawking Solutions}

An interesting set of solutions arises when we consider solutions
for which the base manifold admits a tri-holomorphic Killing vector i.e.
a Killing vector which preserves the
hyper-K\"ahler structure. It has been shown
\cite{gibbons:1988} that such manifolds
are Gibbons-Hawking spaces \cite{Gibbons:1979zt}.
We also make the important assumption,
which leads to much simplification, that the Killing vector generates
a symmetry of the full solution (including all scalars and gauge fields).
The analysis of this section generalises section 3.7 of \cite{classb}.

Locally, we can choose co-ordinates $x^5, x^i$ for $i=1,2,3$ on $M_4$ with
the tri-holomorphic Killing-vector given by $\partial_5$.
The Gibbons-Hawking base metric can then be written as
\be
ds_4{}^2 = H^{-1} (dx^5 + \chi)^2 +H \delta_{ij} dx^j dx^j
\ee
where $\chi=\chi_i dx^i$, and $H$, $\chi$ are independent of $x^5$.
In addition
\be
\nabla \times \bchi = \nabla H
\ee
so in particular $H$ is harmonic on $\bR^3$.
In this section $\nabla$ will
be the gradient and $\nabla^2$ will be the Laplacian on $\bR^3$.

We will find it convenient to introduce the vierbein
\bea
e^5 = H^{-{1 \over 2}} (dx^5 + \chi), \quad e^i = H^{1 \over 2} dx^i
\eea
and positive orientation on the base is defined with respect
to $e^5 \wedge e^1 \wedge e^2 \wedge e^3$.

To proceed we introduce one-forms
\be
\Lambda^I \equiv (\Lambda^I{}_j) dx^j
\ee
so that we can write
\bea
\Theta^I &=& -{1 \over 2} (dx^5+\chi) \wedge
(\Lambda^I{}_j) dx^j
\nn
&-&{1 \over 4} H \epsilon_{ijk}
(\Lambda^I{}_k) dx^i \wedge dx^j
\eea
Closure of $\Theta^I$,  \p{eqn:bianch},
implies that $d \Lambda^I=0$, so that locally $\Lambda^I = d W^I$
for some functions $W^I$. In addition, $d \Theta^I=0$ also implies
\be
\nabla^2 (H W^I)=0
\ee
Hence
\be
W^I = H^{-1} K^I
\ee
where $K^I$ are harmonic functions on $\bR^3$.

Next, we consider the gauge equations ({\ref{eqn:timegauge}});
it is straightforward to see that this is equivalent to
\be
\nabla^2 (f^{-1} X_I) = {1 \over 24} \nabla^2
\big( H^{-1} C_{IPQ} K^P K^Q \big)
\ee
which we solve by taking
\be
\label{eqn:scalsol}
f^{-1} X_I = {1 \over 24} H^{-1} C_{IPQ} K^P K^Q +L_I
\ee
where $L_I$ are some more harmonic functions on $\bR^3$.
Lastly, we shall solve for $\omega$. It is convenient to set
\be\label{omdefn}
\omega = \omega_5 (dx^5 + \chi)+ {\hat{\omega}}
\ee
where
\be
{\hat{\omega}}={\hat{\omega}}_i dx^i.
\ee
Recall that
\be
(d \omega)^+ = -{3 \over 2} f^{-1} X_I \Theta^I
\ee
where ${}^+$ denotes the self-dual projection on the Gibbons-Hawking base.
On using the expression for $f^{-1} X_I$ given in ({\ref{eqn:scalsol}})
we obtain
\bea
\label{eqn:selfom}
\nabla \times {\hat{\omega}} &=& H \nabla \omega_5 -
\omega_5 \nabla H +{3 \over 2}
\big({1 \over 24} C_{IPQ} K^P K^Q
\nn
&+&H L_I \big) \nabla (H^{-1} K^I)
\eea
From the integrability condition of this equation we find the constraint
\bea
\nabla^2 \omega_5 &=& \nabla^2 \big( -{1 \over 48}
 H^{-2} C_{IPQ} K^I K^P K^Q
\nn
 &-&{3 \over 4} H^{-1} L_I K^I
\big)
\eea
which we solve by taking
\be
\label{eqn:omvsol}
\omega_5 =
- {1 \over 48}H^{-2}C_{IPQ} K^I K^P K^Q-{3 \over 4} H^{-1} L_I K^I +M
\ee
where $M$ is another harmonic function on $\bR^3$.
Substituting ({\ref{eqn:omvsol}})
back into ({\ref{eqn:selfom}}) we see that ${\hat{\omega}}$ must satisfy
\be
\label{eqn:hatomeq}
\nabla \times {\hat{\omega}} = H \nabla M - M \nabla H +{3 \over 4}
(L_I  \nabla K^I - K^I \nabla L_I)
\ee
This expression fixes ${\hat{\omega}}$ up to a gradient which can be
removed locally by making a shift in $t$.

The general solution with Gibbons-Hawking base is specified by
$2n+2$ harmonic functions $H$, $K^I$, $L_I$ and $M$ on $\bR^3$.
$H$ determines the Gibbons-Hawking base, and $\omega$ is given by
\p{omdefn}, \p{eqn:omvsol} and \p{eqn:hatomeq}. The scalars $X_I$ and $f$
are determined from \p{eqn:scalsol}. For example, we can multiply
\p{eqn:scalsol} by $X^I$ to get an expression for $f$ and then
substitute this back into \p{eqn:scalsol} to solve for $X^I$.
Finally, the gauge field is determined from \p{eqn:rewritegaug}.

Until now we have not used the condition \p{eqn:jordan}.
If we assume it we obtain some important simplifications.
In particular, the identity \p{symid}
together with ({\ref{eqn:scalsol}}) implies that
\bea\label{niceexpf}
f^{-3} &=& {1 \over 2304} H^{-3} (C_{MNQ} K^M K^N K^Q)^2
\nn
&+&{1 \over 32} H^{-2}(C_{MNQ} K^M K^N K^Q) K^I L_I
\nn
&+&{9 \over 16} H^{-1} C^{IJM} C_{IPQ} K^P K^Q L_J L_M
\nn
&+&{9 \over 2} C^{IJM} L_I L_J L_M
\eea
Observe also that in the full 5-dimensional metric
$g_{55}=f^2[f^{-3}H^{-1}-(\omega_5)^2]$ and using \p{niceexpf}
we obtain
\bea\label{simpeff}
f^{-3} H^{-1}-(\omega_5)^2 &=&
{1 \over 24}H^{-2} C_{MNQ} K^M K^N K^Q M
\nn
&+&{9 \over 16} H^{-2} C^{IJM} C_{IPQ} K^P K^Q L_J L_M
\nn
&+&{3 \over 2} H^{-1} M L_I K^I -{9 \over 16} H^{-2}
(K^I L_I)^2
\nn
&+&{9 \over 2} H^{-1} C^{IJM} L_I L_J L_M -M^2
\eea
which contains no products of more than four of the
harmonic functions $K^I$, $L_I$ in each term.

Henceforth we will take the hyper-K\"ahler base to be $\bR^4$
equipped with metric
\bea
\label{hypkahl}
ds^2 (\bR^4) &=&H\big[dx^i dx^i \big]+H^{-1}(d \psi + \chi_i dx^i)^2
\nn
&=& H(dr^2+r^2 \big[d \theta^2 + \sin^2 (\theta)d \phi^2 \big])
\nn
&+&H^{-1} (d \psi+\cos \theta d \phi)^2
\eea
with $H=1/|{\bf x}|\equiv 1/r$ and we observe
that $\chi_i dx^i = \cos\theta
d\phi$ satisfies $\nabla\times {\bchi}=\nabla H$.  The range of the
angular coordinates are $0<\theta<\pi$, $0<\phi<2 \pi$ and
$0<\psi<4\pi$.

Before proceeding to examine some black ring solutions, it is
useful to compare our conventions with the Gibbons-Hawking
solutions of the minimal theory. These are given in terms of
four harmonic functions $H,K,L$ and $M$, and were presented in
\cite{classb}. In particular, we
note that for the minimal solution, $C_{111}={2 \over \sqrt{3}}$
and $X^1 = \sqrt{3}$, $X_1 = {1 \over \sqrt{3}}$. Moreover, we have
$\Theta^1 = -{2 \over \sqrt{3}} G^+$. Hence, after a
straightforward computation we obtain
\be
\label{eqn:onerel}
K^1 = -2 \sqrt{3} K,\qquad   L_1 = {1 \over \sqrt{3}} L
\ee

\section{Black Ring Solutions}

Our ansatz for the multi-black ring solutions is given by
\bea
\label{eqn:newgensol}
K^I &=& \sum_{i=1}^N q^I{}_i h_i
\nn
L_I &=& \lambda_I  +{1 \over 24} \sum_{i=1}^N (Q_{Ii} -
C_{IJK} q^J{}_i q^K{}_i) h_i
\nn
M &=& \frac{3}{4}  \sum_{i=1}^N \lambda_I q^I{}_i - \frac{3}{4}
 \sum_{i=1}^N \lambda_I q^I{}_i |{\bf{x}}_i|h_i
\eea
where $h_i$ are harmonic functions in $\bR^3$ centred at ${\bf x}_i$,
$h_i=1/|{\bf x}-{\bf x}_i|$, and $Q_{Ii}$, $q^I_i$ and $\lambda_I$ are
constants. To see that this includes the multi
black ring solutions of the minimal theory found in \cite{jgconcentric},
we simply make the identifications
\be
q^1_i={\sqrt 3} q_i,\qquad
Q_{1i}=2 {\sqrt 3} Q_i,\qquad
\lambda_1=\frac{1}{\sqrt 3}
\ee
Note that $i$ labels the $N$ rings. We also note that for the special
case when $N=1$ and ${\bf x}_1=0$, i.e. all of the harmonic functions in
$\bR^3$ are centred at the origin,
that we obtain the general rotating
black hole solution of \cite{Chamseddine:1998yv}.

For simplicity, we will now continue our analysis of these solutions
in the special case that $X^I$ take values in a symmetric space.
In this case we have the identity \p{eqn:jordan}
and we can use the expression for $f$ given in \p{niceexpf}.

To ensure asymptotic flatness, we shall
require that $f \rightarrow 1$ as $r \rightarrow \infty$, and
hence we must have
\be
{9 \over 2} C^{IJM} \lambda_I \lambda_J \lambda_M =1
\ee
We observe that
\be
X_I \rightarrow \lambda_I , \quad X^I \rightarrow \lambda^I\equiv
{9 \over 2} C^{IJK} \lambda_J \lambda_K
\quad {\rm as} \ r \rightarrow \infty \ .
\ee
The sub-leading corrections are easy to calculate
and we find, for example,
\be
X_I=\lambda_I+\frac{1}{24}\left[\mu_I-\lambda_I
(\lambda^J\mu_J)\right]\frac{1}{r}+\dots
\ee
where
\be
\mu_I\equiv \sum_{i=1}^N (Q_{Ii}-C_{IJK} q^J{}_i q^K{}_i)
+ C_{IJK} \sum_{i,j=1}^N q^J{}_i q^K{}_j
\ee
and we see that  the solutions carry scalar charge.
It is also straightforward to calculate the
electric charges carried by the solution. We find that
\be\label{charges}
\frac{1}{2\pi^2}\int_{S^3} Q_{IJ} *F^J=\frac{1}{2}\mu_I \ .
\ee
{}For the ADM mass we find
\be
M_{ADM} = {\pi \over 8 G} \lambda^I \mu_I
\ee
Noting that if we contract \p{charges} with $\lambda^I$ we obtain
$(1/2)\lambda^I\mu_I$, we see that the ADM mass is
consistent with the BPS bound. Explicit expressions
for the angular momentum will be given below for the special
case that all of the harmonic functions
have centres lying on the $z$-axis.

\subsection{Near-Horizon Analysis}

We now analyse what happens as ${\bf x} \rightarrow {\bf x}_i$ for
some fixed $i$. Our analysis essentially
follows that of \cite{jgconcentric}.
We first make a rotation so that ${\bf{x}}_i$ is at
$(0,0,-R_i{}^2/4)$
and set up new spherical polar coordinates
$(\epsilon_i,\theta_i,\phi_i)$ in $\bR^3$ centred on ${\bf x}_i$ and
then consider an expansion in $\epsilon_i$.
After doing this, and solving ({\ref{eqn:hatomeq}}), we find
a coordinate singularity at $\epsilon_i=0$.
To see this, it is useful to note the following expansions:
\bea
\label{eqn:expans}
f &=& {16 \over R_i^2 \nu_i^2} \epsilon_i{}^2 + O(\epsilon_i)^3
\nn
H f^{-1} &=& {\nu_i^2\over 4} \epsilon_i^{-2}
+ \kappa^1{}_i (\epsilon_i)^{-1} + O((\epsilon_i)^0)
\nn
f^2 \omega_5 &=& -{2 \over \nu_i} \epsilon_i
+ \kappa^2{}_i (\epsilon_i)^2 + O(({\epsilon_i})^3)
\nn
f^2 (f^{-3} H^{-1} - (\omega_5)^2) &=& {1 \over 4}
(\ell_i)^2 + \kappa^3{}_i \epsilon_i +  O(({\epsilon_i})^3)
\eea
where $\kappa^1{}_i, \kappa^2{}_i, \kappa^3{}_i$
are constants whose value is not important in
the context of this discussion, and we have set
\be
\label{eqn:constva}
\nu_i = \big( {1 \over 6} C_{IPQ}
q^I{}_i q^P{}_i q^Q{}_i \big)^{1 \over 3}
\ee
and
\bea
\label{eqn:constvb}
\ell_i &=& \nu_i^{-2} \bigg(
{1 \over 16} C^{IJM} C_{IPQ} q^P{}_i q^Q{}_i
(Q_{Ji}-C_{JST} q^S{}_i q^T{}_i)
\nn
&\times& (Q_{Mi}-C_{MUV} q^U{}_i q^V{}_i)
\nn
&-&{1 \over 16} [q^I{}_i (Q_{Ii}-C_{IST} q^S{}_i q^T{}_i)]^2
-3 \nu_i^3 (R_i)^2 \lambda_I q^I{}_i
\bigg)^{1 \over 2}
\nn
\eea
where we have assumed that $\nu_i>0$ and $\ell_i^2>0$.

 Motivated by a similar
analysis in \cite{eemr} we then introduce new coordinates
\bea\label{newcoords}
dt&=&dv+(\frac{b_2}{\epsilon_i^2} +
\frac{b_1}{\epsilon_i})d\epsilon_i
\nn
d\psi&=&d\phi_i'+2 (d\psi'+
\frac{c_1}{\epsilon_i}d\epsilon_i)
\nn
\phi_i&=&\phi_i'
\eea
for constants
$b_j$ and $c_j$.  In order to eliminate a $1/\epsilon_i$ divergence in
$g_{\epsilon_i\psi'}$ and a $1/\epsilon_i^2$ divergence in
$g_{\epsilon_i\epsilon_i}$ we take
\be
b_2 = {\ell_i \nu_i^2 \over 8}  \ , \qquad c_1 = -{\nu_i \over 2 \ell_i}
\ee

A $1/\epsilon_i$ divergence in
$g_{\epsilon_i\epsilon_i}$ can be eliminated by a suitable choice for
$b_1$, whose explicit expression is not illuminating.
The metric can now be
written
\bea
&& ds^2 =- {256 \over R_i^4 \nu_i^4} \epsilon_i^4 dv^2 -
\frac{4}{\ell_i} dv d\epsilon_i +
{32 \over R_i^4 \nu_i} \sin^2 \theta_i \epsilon_i^3 dv d\phi_i'
\nn
&+& {8 \over \nu_i} \epsilon_i dv d\psi'
+ \ell_i^2 d{\psi'}^2 + {\nu_i^2 \over 4} \left[d\theta_i^2
+ \sin^2 \theta_i d\phi_i'^2\right]
\nn
&+&2g_{\epsilon_i \phi_i'} d\epsilon_i d\phi_i'
+ 2g_{\epsilon_i\psi'}d\epsilon_i d\psi'
+g_{\epsilon_i\epsilon_i}d\epsilon_i^2
\nn
&+&2g_{\psi'\phi_i'}d\psi' d\phi_i'
 +2 g_{v \theta_i} dv d \theta_i
+2 g_{\psi' \theta_i} d \psi' d \theta_i
\nn
&+& 2 g_{\epsilon_i \theta_i} d \epsilon_i d \theta_i
+ 2 g_{\theta_i \phi_i'} d \theta_i d \phi_i'
+ \ldots
\eea
where
$g_{\epsilon_i\psi'}$ and $g_{\epsilon_i\epsilon_i}$
are ${\cal O}(\epsilon_i^0)$; $g_{\psi'\phi_i'}$ and $g_{\epsilon_i
\theta_i}$ are ${\cal
O}(\epsilon_i)$; $g_{v \theta_i}$ is ${\cal O}(\epsilon_i^5)$;
$g_{\psi' \theta_i}$ is ${\cal O}(\epsilon_i^2)$; and
$g_{\theta_i \phi_i'}$ is ${\cal O}(\epsilon_i^4)$
whose explicit forms are unimportant for our considerations here, and
the ellipsis denotes terms involving sub-leading (integer) powers of
$\epsilon_i$ in all of the metric components explicitly indicated.

The determinant of this metric is analytic in $\epsilon_i$. At $
\epsilon_i=0$ it
vanishes if and only if $\sin^2 \theta_i = 0$, which just
corresponds to
coordinate singularities.
It follows that the inverse metric is also
analytic in $\epsilon_i$ and hence the above
coordinates define an analytic
extension of our solution through the surface $\epsilon_i=0$.

The supersymmetric Killing vector field $V = \partial_v$ is
null at $\epsilon_i=0$. Furthermore $V_\mu dx^\mu = -(2/\ell_i)
d\epsilon_i$ at
$\epsilon_i=0$, so $V$
is normal to the surface $\epsilon_i=0$. Hence $\epsilon_i=0$
is a null hypersurface and
a Killing horizon of $V$, i.e., the black
ring has an event horizon which is
the union of the Killing Horizons for each $\epsilon_i=0$.
Furthermore, by expanding out the determinant of
the metric obtained by restricting to the surface
on which $v$, $\theta_i$ and $\epsilon_i$ are constant,
it is straightforward to show that there are no closed
time-like curves (CTCs) at the horizon.

In the near horizon limit defined by scaling $v\to v/\delta$,
$\epsilon_i\to \delta\epsilon_i$ and then taking the limit
$\delta\to 0$, we find that
the metric is locally the product of $AdS_3$ with radius $\nu_i$ and a
two-sphere of radius ${\nu_i \over 2}$.

We can read off the geometry of a spatial cross-section of the
horizon:
\be
ds^2_\mathrm{horizon} =  \ell_i^2 d{\psi'}^2 + {\nu_i^2 \over 4}
\left[d\theta_i^2
+ \sin^2 \theta_i d\phi_i'^2\right] \ .
\ee
We see that the horizon has geometry $S^1 \times S^2$, where the $S^1$
and round $S^2$ have radii $\ell_i$
and ${\nu_i \over 2}$, respectively. This is
precisely the geometry of the event horizon for a single
supersymmetric black ring. The area of this specific horizon is given by
\be
{\cal{A}} = 2 \pi^2 \ell_i \nu_i^2 \ .
\ee
Note that since $\ell_i$ depends on $\lambda_I$,
the black ring horizon area depends on the values taken
by the scalars at asymptotic infinity. We will discuss how this
is related to the angular momentum in the next sub-section.

We also note that near the pole
\bea
X_I &=& {1 \over 6} \nu_i^{-2} C_{IST} q^S{}_i q^T{}_i
+ O(\epsilon_i)
\nn
X^I &=& \nu_i^{-1} q^I{}_i + O(\epsilon_i)
\eea
so the scalars are regular near the horizon. Moreover,
although the $\Theta^I$ are
not regular at the horizon, it is straightforward
to show, using the expansions given in
({\ref{eqn:expans}}), that the gauge field
strengths $F^I$ are also regular at the horizon.

The $i$th ring has dipole charges defined by
\be
D^I{}_i = {1 \over 16 \pi G} \int_{S^2_i} F^I = {q^I{}_i \over 8G}
\ee
where $S^2_i$ encloses the $i$th black ring only once,
and can be taken to be the $S^2$ at the $i$th horizon.

The $S^1$ direction of the rings all
lie on an orbit of the tri-holomorphic
Killing vector $\partial_5$ and hence describe concentric rings (see
\cite{jgconcentric} for further comments). We also note that if we
set ${\bf x}_i=0$ for one value of $i$ we obtain a general rotating
black hole with topology $S^3$ of the kind found
in \cite{Chamseddine:1998yv}, sitting at the centre of the rings.

\subsection{Poles on the $z$-axis.} We now consider the solutions
with all poles located along the $z$-axis, where we can analyse the
solutions in more detail.
Consider the general solution ({\ref{eqn:newgensol}})
with ${\bf x_i}=(0,0,-k_iR_i^2/4)$ and $k_i=\pm 1$. Thus
\be
h_i=(r^2+\frac{k_iR_i^2}{2}r\cos\theta+\frac{R^4_i}{16})^{-1/2} \ .
\ee
We can solve ({\ref{eqn:hatomeq}}) with $\hat\bomega$
only having a non-zero $\phi$ component, $\hat\omega_\phi$, that is
a function of $r$ and $\theta$ only. In particular,
in addition to $\partial_t$ and $\partial_5$,
these solutions have an extra $U(1)$
symmetry generated by $\partial_\phi$.

To solve ({\ref{eqn:hatomeq}}) we write
$\hat\omega=\hat\omega^L+\hat\omega^Q$ where $\hat\omega^L$ is linear
in the charges $q_i$ and independent of $Q$
and $\hat\omega^Q$ contains the dependence on $Q$.
We find
\be\label{linom}
\hat\omega^{L}=-\sum_{i=1}^{N}\frac{3 q^I{}_i \lambda_I}{4}
[1-(r+\frac{R_i^2}{4})h_i](\cos\theta+k_i)d\phi
\ee
and
\bea
\hat\omega^{Q}=&-&\frac{1}{256}\sum_{i<j}
\frac{1}{(k_iR_i^2-k_jR_j^2)}
\nn
&\times&
\big[
q^I{}_i (Q_{Ij}-C_{IJK} q^J{}_j q^K{}_j)
\nn
&-& q^I{}_j (Q_{Ii}-C_{IJK} q^J{}_i q^K{}_i) \big]
\nn
&\times&  h_i h_j \left[\frac{16}{h_i^{2}}+\frac{16}{h_j^{2}}
-\frac{32}{h_ih_j}-(k_iR^2_i-k_jR_j^2)^2\right] d \phi
\nn
\eea

By considering the asymptotic form of the solution we find that
the angular momentum is given by
\bea
\label{diffjays}
J_{1}&=& {\pi \over 48 G} \big[2\sum_{i,j,m=1}^N
C_{IPQ} q^I{}_i q^P{}_j q^Q{}_m
\nn
&+&3\sum_{i,j=1}^N q^I{}_i (Q_{Ij}-C_{IJP} q^J{}_j q^P{}_j)\big]
\nn
&-&{3\pi \over 8G} \sum_{i=1}^N j_i(k_i-1)
\nn
J_2&=&J_1+\frac{3\pi}{4G}\sum_{i=1}^{N}j_ik_i \ .
\eea
where we have defined
\be
j_i\equiv q^I_i\lambda_IR^2_i
\ee

If we have a single black ring with $N=1$ we now see that the
moduli dependence of the area of the event horizon appearing
in the expression for $\ell_1$ in \p{eqn:constvb} can be re-expressed in
terms of the conserved charge $J_2-J_1$,
consistent with \cite{larsen:1995}.
More generally, for the multi-ring solutions,
with poles on the $z$-axis, the moduli dependence in $\ell_i$ can be
expressed in terms of $j_i$, which, from \p{diffjays}, have the natural
interpretation as fixing the contribution
to $J_2-J_1$ coming from the $i$th ring. It would be interesting to
check that this also holds for poles not all on the $z$-axis.
The black hole entropy also depends on $Q_I$ which are quantised
electric charges when the model comes from,  for example,
the reduction of D=11 supergravity on a Calabi-Yau three-fold
(e.g. \cite{Chou:1997ba}).
It also depends on the $q^I$, which we saw above
are dipole charges, which
are expected to be quantised, similarly.

We would also like to check whether there are any Dirac-Misner strings
that might require making periodic identifications of the time coordinate.
From the reasoning given for the computation of black ring
solutions in \cite{jgconcentric},  we
demand that $\hat\omega=0$ at $\theta=0,\pi$.
The expression for $\hat\omega^{L}$ in \p{linom} satisfies
these conditions. In order for the same to hold for
$\hat\omega^{Q}$ we require that
\be
q^I{}_i (Q_{Ij}-C_{IJK} q^J{}_j q^K{}_j)= q^I{}_j
(Q_{Ii}-C_{IJK} q^J{}_i q^K{}_i)
\ee
for $i \neq j$. A condition which is sufficient
(though not generally necessary)
for this to hold is
\be
\label{eqn:simpcons}
Q_{Ii}-C_{IJK} q^J{}_i q^K{}_i = \Lambda q_{Ii}
\ee
for all $I$, $i$ where $\Lambda$ is constant
and $q_{Ii} \equiv \delta_{IJ} q^J{}_i$.
However, in general, this condition is excessively
constraining on the parameters of the solution.
If we do impose this condition we have
$L_I = \lambda_I + {\Lambda \over 24} \delta_{IJ} K^J$.

\section{Three-Charge Solutions}

In the special case of the 3-charge $STU$-model
with $C_{123}=1$ we obtain
some useful simplifications. In particular we find
\bea
f^{-3} &=& {1 \over 64} (12 L_1 +H^{-1} K^2 K^3)
(12 L_2 +H^{-1} K^1 K^3)
\nn
&\times&
 (12 L_3 +H^{-1} K^1 K^2)
\eea
and so we find
\bea
\label{eqn:sscal}
X_1 &=&\frac{1}{3}{(12L_1 +H^{-1} K^2 K^3)^{2/3} \over
(12 L_2 +H^{-1} K^1 K^3)^{1/3}(12 L_3 +H^{-1} K^1 K^2)^{1/3}}
\nn
X_2 &=&{1 \over 3} {(12L_2 +H^{-1} K^1 K^3)^{2/3} \over
(12 L_1 +H^{-1} K^2 K^3)^{1/3}(12 L_3 +H^{-1} K^1 K^2)^{1/3}}
\nn
X_3 &=&{1 \over 3} {(12L_3 +H^{-1} K^1 K^2)^{2/3} \over
(12 L_1 +H^{-1} K^2 K^3)^{1/3}(12 L_2 +H^{-1} K^1 K^3)^{1/3}}
\nn
\eea
Furthermore:
\be
\nu_i = (q^1{}_i q^2{}_i q^3{}_i)^{1 \over 3}
\ee
and
\bea
\label{eqn:radiusa}
\ell_i &=& (q^1{}_i q^2{}_i q^3{}_i)^{-{2 \over 3}}
\big[ -{1 \over 16} (q^1{}_i)^2 (Q_{1i}-2 q^2{}_i q^3{}_i)^2
\nn
 &-&{1 \over 16} (q^2{}_i)^2 (Q_{2i}-2 q^1{}_i q^3{}_i)^2
 \nn
&-&{1 \over 16} (q^3{}_i)^2 (Q_{3i}-2 q^1{}_i q^2{}_i)^2
\nn
&+&{1 \over 8} q^1{}_i q^2{}_i (Q_{1i}-2 q^2{}_i q^3{}_i)
(Q_{2i}-2 q^1{}_i q^3{}_i)
\nn
&+&{1 \over 8} q^1{}_i q^3{}_i (Q_{1i}-2 q^2{}_i q^3{}_i)
(Q_{3i}-2 q^1{}_i q^2{}_i)
\nn
&+&{1 \over 8} q^2{}_i q^3{}_i (Q_{2i}-2 q^1{}_i q^3{}_i)
(Q_{3i}-2 q^1{}_i q^2{}_i)
\nn
&-&3 q^1{}_i q^2{}_i q^3{}_i R_i^2 (\lambda_1 q^1{}_i +
\lambda_2 q^2{}_i + \lambda_3 q^3{}_i)
\big]^{1 \over 2}
\eea

If we impose the constraint ({\ref{eqn:simpcons}}) which removes
the string singularities, this simplifies to
\bea
\ell_i &=&  (q^1{}_i q^2{}_i q^3{}_i)^{-{2 \over 3}} \big[
-3 R_i^2 q^1{}_i q^2{}_i q^3{}_i \lambda_I q^I{}_i
\nn
&+&{\Lambda^2 \over 16}(q^1{}_i + q^2{}_i + q^3{}_i)
(q^1{}_i + q^2{}_i - q^3{}_i)
\nn
&\times&
(q^1{}_i - q^2{}_i + q^3{}_i) (-q^1{}_i + q^2{}_i + q^3{}_i)
\big]^{1 \over 2}
\nn
\eea
{}For this case, in order to ensure that $f$
is positive, it suffices to take
$\lambda_I>0$, $q^I{}_i>0$, and $\Lambda>0$.
In order to get black rings
with non-vanishing horizon area we also require $\ell_i>0$. We will
comment momentarily on the significance of $\ell_i=0$.

\section{Conclusion}

In this paper we have constructed
new supersymmetric solutions corresponding
to concentric black rings carrying multiple charges. We have shown that
the rings have regular horizons, and have computed the
conserved charges associated with these rings. We have shown
that there are no closed time-like curves at the horizon.
Although we think it is unlikely
that there are any CTCs elsewhere, this remains
to be proven; in particular, it would be
interesting to be able to examine the global causal structure
of the most general solutions in which the poles
of the harmonic functions are not all co-linear.
The main obstacle to this would appear
to be the complicated nature of the solution to ({\ref{eqn:hatomeq}}).

The circumference of the $i$th black ring horizon in our solutions is
given by $2\pi \ell_i$, where $\ell_i$ is defined in \p{eqn:constvb}.
It is interesting that in the special case of a single three charge
ring, it has been argued in \cite{warner:2004} that the geometry
with $\ell_i=0$ corresponds to a regular three-charge supertube
\cite{Mateos:2001qs,Bena:2004wt} without horizon.
It would be interesting to prove that
such geometries are indeed regular and free
from CTCs, since our solutions would then
also include superpositions of
concentric multi-charge regular supertubes
with black rings and a black hole at the common centre.

The detailed analysis of the black ring solutions presented here
focussed on the case in which the scalars take values in
a symmetric space. The reason for this is that for the most
generic type of scalar manifold, one cannot obtain an explicit
expression for the scalar $f$. So, for these more general solutions,
additional assumptions concerning the regularity of the scalars at
asymptotic infinity and at the horizon could be needed in order
to compute the conserved charges and to analyse the near-horizon geometry.
Nevertheless,
it should be quite straightforward to find some black ring solutions
for other types of simple scalar manifold- such as the ``flop transition''
solution examined in the context of
five-dimensional black holes in \cite{sabratransit:1998}.
Moreover, one could also investigate whether even more
general black ring solutions could be found for which the
hyper-K\"ahler base is no longer flat. Finally,
it is an interesting open question
as to whether or not there are
supersymmetric black ring solutions in gauged supergravity theories.

\begin{acknowledgments}
We thank Harvey Reall for a helpful correspondence concerning the moduli
dependence of the area of the black ring horizons.
J.B.G. thanks EPSRC for support.

{\bf Note added:} After this work was completed an interesting paper
appeared \cite{warner:2004} which has some overlap with
section 5 of this paper.
In particular, in the three-charge case, a solution
describing a single black ring with a possible
black hole at the centre is explicitly constructed in  \cite{warner:2004}.
However, the solutions we present for this
model are more general in that we can have an arbitrary number of
concentric rings and we also allow for the possibility of arbitrary
asymptotic values of the scalar fields.
\end{acknowledgments}

%%%%%%%%%%%%%%%%%%%%

\end{document}